\documentclass{nature}

\usepackage{graphicx}
\makeatletter
\let\saved@includegraphics\includegraphics
\AtBeginDocument{\let\includegraphics\saved@includegraphics}
\renewenvironment*{figure}{\@float{figure}}{\end@float}
\makeatother

\usepackage{color}
\usepackage{amsmath,amssymb}
\usepackage[normalem]{ulem}

\usepackage{times}

\title{\textcolor{black}{Activity-controlled Annealing of Colloidal Monolayers}}

\author
{Sophie Ramananarivo$^{1}$, Etienne Ducrot$^{2}$,  Jeremie Palacci$^{1\ast}$\\\
\normalsize{$^{1}$Department of Physics, University of California San Diego, USA}\\
\normalsize{$^{2}$Department of Physics, New York University, USA}}

\begin{document}



\maketitle

\begin{abstract}
\textcolor{black}{Molecular motors are essential to the living, they generate additional fluctuations that boost transport and assist assembly. Self-propelled colloids, that consume energy to move, hold similar potential  for the man-made assembly of microparticles.} \textcolor{black}{Yet, experiments showing their use as a powerhouse in materials science lack. Our work explores the design of man-made materials controlled by fluctuations, arising from the internal forces generated by active colloids.  Here we show a massive acceleration of the annealing of a monolayer of passive beads by moderate addition of self-propelled microparticles. We rationalize our observations with a model of collisions that drive active fluctuations to overcome kinetic barriers and activate the annealing. The experiment is quantitatively compared with Brownian dynamic simulations that further unveil a dynamical transition in the mechanism of annealing. Active dopants travel uniformly in the system or co-localize at the grain boundaries as a result of the persistence of their motion. Our findings uncover the potential of  man-made materials controlled by internal activity and lay the groundwork for the rise of materials science beyond equilibrium.}
\end{abstract}

\textcolor{black}{In the classical picture of Brownian motion, particles and fluid molecules are passive and driven by thermal fluctuations\cite{Jean-perrin, Brown:1828jv, Einstein:1905ty}. \textcolor{black}{The incessant motion of suspended particles is then due to multiple collisions with the surrounding fluid molecules}. In living systems, molecular motors ballistically zoom around, push and pull on each others, and exert forces on passive elements, yielding a riotous environment radically different from \textcolor{black}{the more mundane equilibrium one}. This non-equilibrium dynamics induces additional fluctuations\cite{Brugues:2014vc,Wittmann:2001ed} that boost transport in biological systems\cite{Brangwynne:2008dx, Guo:2014dw} or quickly bring building blocks into contact while reducing unwanted connections\cite{Hess:2006fq}. \textcolor{black}{Furthermore, the effect of internal agitation was explored in the different context of driven vortex lattices in high-temperature superconductors\cite{Vinokur:2011ub,Paltiel:2000gd}}. Active colloids, which were made available by recent synthetic progress, similarly inject energy locally and drive systems out of equilibrium\cite{Sengupta:2012iea}.}
\textcolor{black}{Developments on active colloids have surged in recent years, focused on exotic behavior without equilibrium counterpart: broken microscopic reversibility\cite{di2010bacterial,sokolov2010}, \textcolor{black}{and mesoscopic fluxes such as} flocks\cite{bricard2013}, flows\cite{Wu:2017ct,Lushi:2014fn}, phase separation\cite{Cates:2015ft,theurkauff2012,buttinoni2013,palacci2013} or anomalous transport of dispersed tracers\cite{wu2000,leptos2009,mino2011}. It largely overlooked their practical impact to deploy materials science beyond equilibrium and control matter. Active particles added to a material generate forces from within and provide a unique opportunity to overcome the naturally occurring kinetic barriers and modulate the energy landscape of soft materials. This potential was highlighted by Reichhardt \& Reichhardt\cite{Reichhardt:2004dg}, \textcolor{black}{who} numerically studied the case of a single colloid driven through a colloidal crystal. They showed that a local melting can occur when the electrostatic charge of the driven particle is larger or comparable to that of the colloids comprising the crystal, \textcolor{black}{leading to} the generation of lattice defects that increase the drag force and generate large noise fluctuations\cite{Reichhardt:2004dg}.  Opportunities in colloidal assembly were further stressed by recent numerical works that considered mixtures of active, autonomously-driven particles, in passive phases. They showed that the presence of active particles can favor the annealing of \textcolor{black}{a passive polycrystal by melting its grain boundaries}\cite{van2016,van2017,kummel2015} or act as an internal field to control the structural and dynamical \textcolor{black}{properties} of a colloidal gel\cite{KOmar:2018ho}. Notwithstanding, experimental realizations of mixtures  of active particles in a passive phase are scarce and \textcolor{black}{have not demonstrated} control of the passive phase. Dietrich {\it et al}\cite{Dietrich:2018hp} showed that the motion of self-propelled particles is altered by the presence of a loosely packed crystal, an interesting result,  however solely focused on the dynamics of the active individuals. Alternatively, Kummel {\it et al}\cite{kummel2015} showed that \textcolor{black}{in sparse colloidal layers, active particles tend to gather colloids into dynamical clusters, while in denser ones, they accumulate at the grain boundaries separating crystalline domains, where they initiate melting and widen those disordered regions}. \textcolor{black}{Although further numerical modeling of their system suggests that this localized melting could lead to a large defect-free crystal\cite{kummel2015}, the exploitation of active dopants as a workforce for material design still remains to be achieved experimentally.} Experimental roadblocks thwarted progress towards this goal, \textcolor{black}{preventing the implementation of this seemingly simple new pathway for material control.} In this work, we devised a  microfluidic system of confining arenas with slanted walls, that allows exchange of active particles with an external reservoir and enables a uniform injection of energy to the passive phase. It made possible the control of the annealing of a colloidal monolayer in time and space by harnessing the activity of embedded active intruders, offering the first experimental demonstration of man-made materials externally controlled by active noise.}

\begin{figure}
\centering
\includegraphics[width=0.9\linewidth]{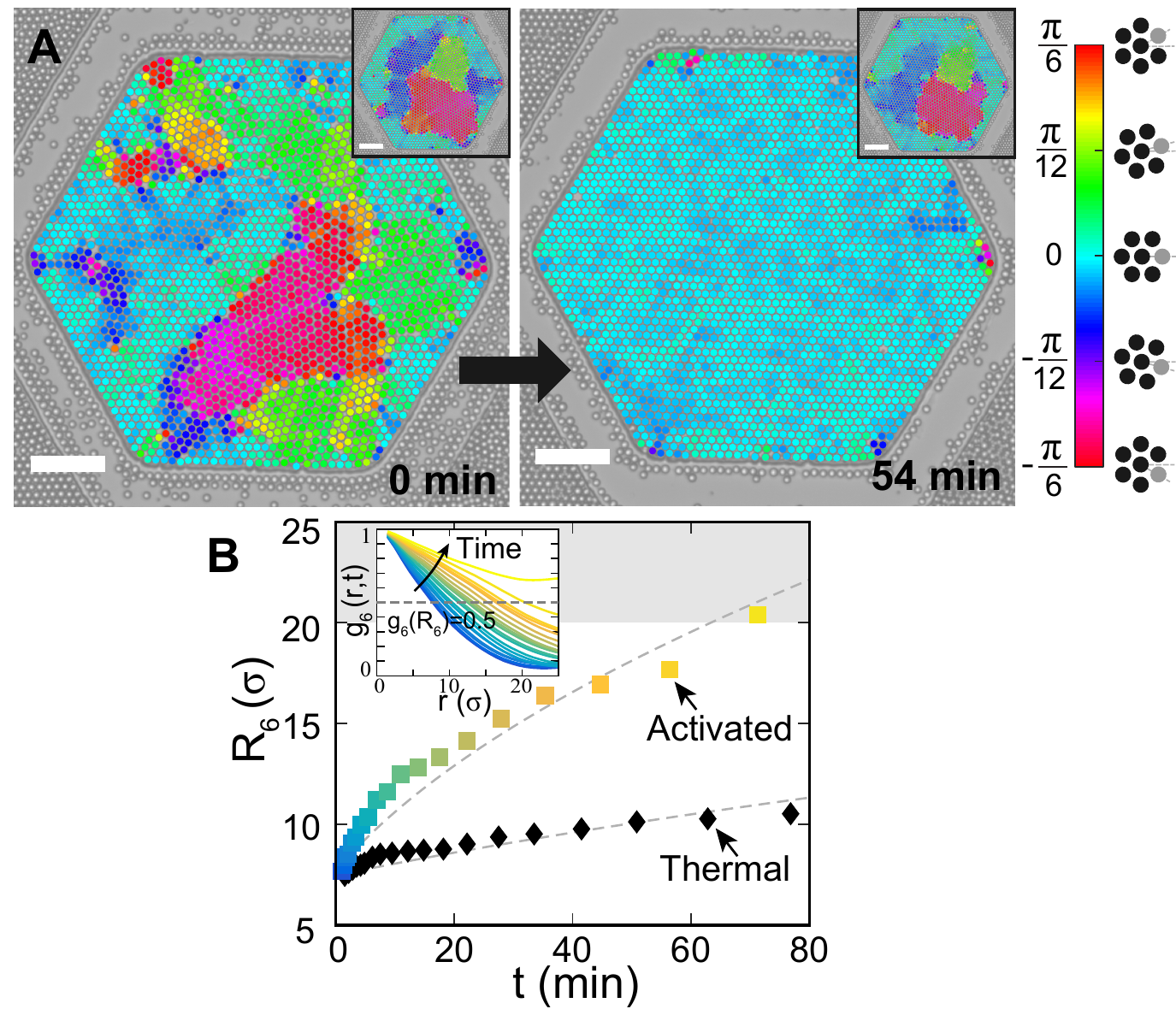}
\centering
\linespread{1}
\caption{{\bf Accelerated annealing}. {\bf (A)} Optical microscopy of a hexatic monolayer of passive beads (circles) containing a small fraction of active intruders (not visible at this magnification). The passive beads are color-coded by the local orientational field of the hexatic order. The system is initially quenched in adjacent polycristallites (left panel) and  rapidly relaxes towards an ordered monolayer (right panel), following the activation of the self-propelled particles in the layer. The thermal experiment, in the absence of active particles, is significantly slower  and shows  little evolution within the same time span [insets]. Scale bars, $50 \mu m$. {\bf (B-inset)} Bond orientational correlation function $g_6(r,t)$ as a function of distance $r$ and for increasing times $t$  (blue to yellow color gradient) [see main text]. It increases with time as long-range order develops: large grains grow at the expense of smaller ones. {\bf (B)}:  Time-evolution of the characteristic grain radius $R_6$,  in units of colloid diameters $\sigma$, defined from  $g_6(R_6,t)=0.5$ for thermal (black) monolayer or in the presence of embedded self-propelled particles (color, $V=10\mu m/s$, $\alpha=1.3\%$) and fit by a normal grain growth  $R_6(t)^2=R_6(0)^2+\gamma t$ (dotted lines, see Main text). The grey area delimits grain sizes $>20 \sigma$ which cannot be measured accurately in the experiment. \label{Fig1}}
\end{figure}

\textbf{Accelerated annealing.} Our experimental system consists of a suspension of beads of diameter $\sigma=5 \mu m$ (Sigma aldrich, silicon dioxide, 44054) filling a hexagonal well which is embossed on the bottom substrate of a microfluidic chamber. After sedimentation, the beads form a dense monolayer of spheres at a surface fraction $\Phi_s\sim0.68\pm0.03$ with a local hexatic order [Methods and SI-Fig.1]. The rapid increase of the surface density of the sedimenting particles quenches the system into adjacent crystallites separated by grain boundaries [Fig.\ref{Fig1}A],  a metastable state that eventually converts into a single crystal. At thermal equilibrium, it takes about $12$ hours for a system of 2200 beads to form a single crystal in a hexagonal chamber of width $260\mu m$.  We explore the impact of the addition of a small fraction of active intruders to this initially  polycrystalline monolayer of silica spheres. The active intruders are light-activated microswimmers, consisting of a hematite (Fe$_2$O$_3$) portion protruding out of a polymer sphere\cite{palacci2013}; they are $2\mu m$ in diameter and exhibit a two-dimensional persistent random walk in free space\cite{howse2007}.  In a nutshell, the iron oxide photocatalytically  decomposes a solution of hydrogen peroxide fuel  after activation by a wavelength $\lambda=390-480nm$, resulting in subsequent propulsion of the particle in the concentration gradient\cite{palacci2013}. The velocity is controlled by the intensity of the light\cite{palacci2013} that illuminates the microfluidic chamber uniformly. We periodically interrupt the propulsion of the active particles by short periods of extinction, light off, to allow the swimmers possibly wedged between the passive spheres and the substrate to reorient and be released. We observe a qualitative difference of dynamics between thermal and active annealing.  In the latter, streaks form  following the displacement of the intruders in the crystal  [MovieS1].  Within 40 mins, the activated system reorganizes towards a fully ordered state [Fig.\ref{Fig1}A] whereas its thermal counterpart -with no swimming activity- has undergone little change [Fig.\ref{Fig1}A-insets, MovieS1].

To gain insight into the phenomenon, we vary the numeric fraction $\alpha$ of intruders in the monolayer, $\alpha=0.1-5\%$ of the total number of passive beads,  and their propulsion velocity $V$ using the control offered by light.  We  perform the experiments in larger systems  with $5600$ passive beads, for which the reorganization can be followed on larger spatial and temporal  scales. The passive beads are gravitationally confined in a hexagonal arena, whose slant and limited depth $h\sim 4\mu m$,  allow the self-propelled particles to  escape and enter\cite{koumakis2013}. This avoids the accumulation of the active particles at the walls and maintains a constant surface fraction $\Phi_s$ of passive beads and swimmers, an important prescription to the present study of annealing. 
 
We quantify the temporal and spatial properties of the monolayer using the  bond-orientational correlation function $g_6(r,t)$, calculated from the six-fold bond-order parameter [see Methods, SI-Fig.1]. The function $g_6(r)$ increases with time as long-range order develops [Fig.\ref{Fig1}B-inset]:  the system coarsens and large grains develop at the expense of the smaller ones  with the walls favoring crystallites that are geometrically compatible with the boundaries [MovieS2]. A characteristic grain size $R_6$ is set by the threshold of the bond-orientational correlation function, $g_6(R_6,t)=0.5$ [Fig.\ref{Fig1}B] as previously introduced in thermal systems\cite{lavergne2017,sicilia2008}, and  confirmed as a relevant length scale of the grain structure by the time-independent rescaling $g_6(r,t)=f(r/R_6(t))$ [SI-Fig.2].  The grain size $R_6(t)$ increases with time and shows a reasonable agreement  with the normal law  for a curvature-driven grain growth, $dR_6/dt=\gamma/R_6$, where $\gamma$ is the mobility of the grain boundary\cite{rollett2004} [Fig.\ref{Fig1}B]. \textcolor{black}{It offers a convenient proxy to quantify the large speeding up of the observed annealing [see SI].  Devising more intricate models, that capture in finer details the coarsening mechanisms in such strongly out-of-equilibrium systems, constitutes exciting developments but are beyond the scope of this work.} We further use $\gamma$ as a measure of the observed dynamics and study its dependence on the speed and the fraction of active intruders. We report a massive increase in $\gamma$, up to 40-fold, compared to a thermal system. 

\begin{figure}
\includegraphics[width=\linewidth]{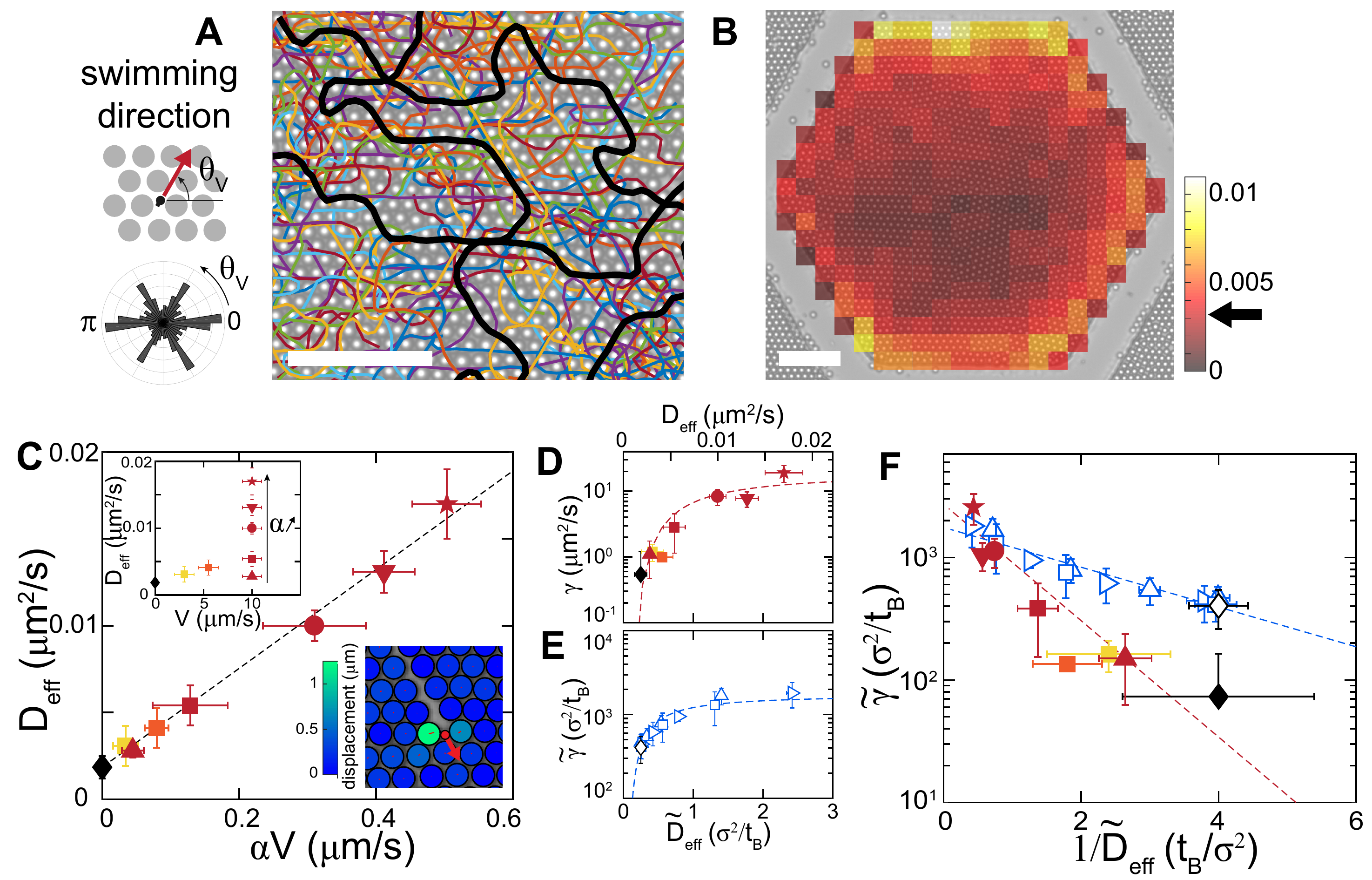}
\centering
\linespread{0.95}\vspace{-1cm}
\caption{{\bf Microscopic dynamics.} {\bf (A)} Individual trajectories of self-propelled particles in an ordered monolayer. Some trajectories are highlighted in black for better visibility. The probability distribution of propulsion direction  $\theta_V$  of the active particles shows peaks along the natural directions of the crystal. {\bf (B)} Probability distribution of the presence of swimmers in the colloidal layer over the first 20min of an experiment. It is uniform (value for a uniform distribution indicated by an arrow), the swimmers do not localize on the walls, the grain boundaries or elsewhere. Scale bars, $50 \mu m$. {\bf (C)} {\bf [top inset]} Measured diffusivity $D_\text{eff}$ of the passive colloids of the monolayer as a function of the speed $V$ of the active particles. Colors distinguish speeds $V=3$ (yellow), $5$ (orange), $10 \mu m/s$ (dark red); and symbols refer to different fraction of swimmers $\alpha=0.4$ (upward triangle), $1.3$ (square), $3.1$ (circle), $4.1$ (downward triangle), and $5.0\%$ (star). The data collapse onto a master curve $D_{\text{eff}}=D^*+\beta \alpha V$ when plotted as a function of $\alpha V$, with $D^*$ the diffusivity for a purely passive system (black diamond) simply described by a model of collisions of the passive beads with the active particles (see Main text). {\bf (C)[bottom inset]} Displacement of colloids induced by a swimmer (red dot, the arrow indicates the direction of motion); the color-coding of particles denotes the amplitude of displacement (traced by red lines) following a collision. {\bf (D)} Grain mobility $\gamma$ as a function of $D_\text{eff}$ in experiments; the presence of intruders speed up the reorganization of the monolayer by nearly two orders of magnitude. {\bf (E)} $\gamma(D_\text{eff})$ obtained in simulations by varying the velocity of self-propelled particles at a fixed persistence time comparable with the experiments [see Methods]. Data are nondimensionalized using $\sigma$ and the Brownian time $t_B$ as a characteristic length and time (see Main text); symbols differentiate between fractions of swimmers $\alpha=0.3$ (upward triangle), $0.6$ (right-pointing triangle), and $1.2\%$ (square).  {\bf (F)} Log-lin plot of $\gamma$ as a function of $1/D_\text{eff}$ shows an Arrhenius-like behavior in the experiment (color symbols) and the  simulations  (hollow symbols).\label{Fig2} }
\end{figure}

\textbf{Microscopic dynamics and activated process.} In order to connect the observed phenomenon with the individual dynamics of the particles, we follow the trajectories of the self-propelled particles in the well. They travel along the local directions of the surrounding crystal with a speed  $V=3-10\mu m/s$, controlled by light, and a persistence time of $\tau_r\sim 2$s comparable with the  values measured in free space [Fig.\ref{Fig2}A, SI-Fig.4]. The trajectories show no apparent directional persistence from one crystallite to the next, as they tend to reorient upon crossing a grain boundary. Swimmers travel uniformly in the chamber with limited accumulation at the walls and show no significant localization at the grain boundaries or elsewhere [Fig.\ref{Fig2}B]. This confirms the absence of Motility-Induced Phase Separation\cite{mccandlish2012,Cates:2010um} that would notably arise from a non-uniform illumination. \textcolor{black}{This result contrasts with previous numerical reports of intruders accumulating at the grain boundaries and initiating the melting from the surface\cite{van2016,van2017}}.

We further quantify the impact of the activity by tracking the individual trajectories of the passive beads. The dynamics are complex and depend on local rearrangements and instantaneous interactions with the self-propelled particles. In virtue of the uniform distribution of the self-propelled particles [Fig.2B], we adopt a coarse-grained approach and consider an ensemble average of the passive particles  during the first 20mins of the experiment. The dynamics are diffusive at long times and we extract the effective diffusion coefficient $D_\mathrm{eff}$ from the linear fit of the mean square displacement [SI-Fig.3].  Experimental results for varying speeds and fraction of swimmers [Fig.2C-top inset] gather on a master curve: $D_\mathrm{eff}=D^* + \beta\ \alpha V$, where $D^*$ is the diffusivity at equilibrium measured independently and  $\beta\sim 3.10^{-2}\mu m$, the single fitting parameter for the experimental data [Fig.\ref{Fig2}C]. We rationalize this result by a kinetic model of intruders colliding with the monolayer, similar in spirit to Mino {\it et al}\cite{mino2011} in the opposite limit of dilute tracers in an active bath. Considering the passive beads as fixed targets, their decorrelation $\tau$ time is set by the inverse of the collision rate with the active particles, $\tau^{-1}\propto\phi_a V\ell$, where $\phi_a\sim\alpha \Phi_S/\sigma^2$ is the number of intruders per unit area,  and $\ell$ is the interaction length transverse to the motion of the swimmers. Each collision induces a displacement $d$, independent of the velocity $V$ at low Reynolds number, so that the active contribution to the diffusivity of the beads scales as  $d^2/\tau\propto \frac{d^2\ell \Phi_S}{\sigma^2} \alpha V$. Comparison with the experiment gives $\frac{d^2\ell \Phi_S}{\sigma^2}\sim 3.10^{-2}\mu m$. This leads to ${d^2\ell}\sim 1 \mu m^3$, that is $d\sim\ell\sim\sigma/5$, in line with experimental observations and a simple scenario dominated by hard-sphere collisions without entrainment [Fig.2C-bottom inset].

Next, we connect the grain mobility $\gamma$ with the behavior of the passive beads in the monolayer. The grain-growth increase correlates with the enhancement of the diffusivity [Fig.\ref{Fig2}D] and follows an Arrhenius-like law, $\gamma\propto \exp{-A/D_\mathrm{eff}}$ [Fig.\ref{Fig2}F]. Active fluctuations that originate from the presence of the self-propelled particles allow to cross the energy barrier $E$ required to displace grain boundaries. We extract from the data, $A\sim E/\mu^*\sim 8.10^{-3}\pm 1.10^{-3}\mu m^2/s$, where $\mu^*$ is the equilibrium mobility relating equilibrium fluctuation and dissipation\cite{barrat2003}, $D^*=kT/\mu^*$,  larger than the Stokes mobility due to additional dissipation  from confinement. It gives $E\sim 4\pm1 \ k T$, a reasonable value for the entropic barrier of a dense monolayer of hard spheres. 

\begin{figure}
\includegraphics[width=0.6\linewidth]{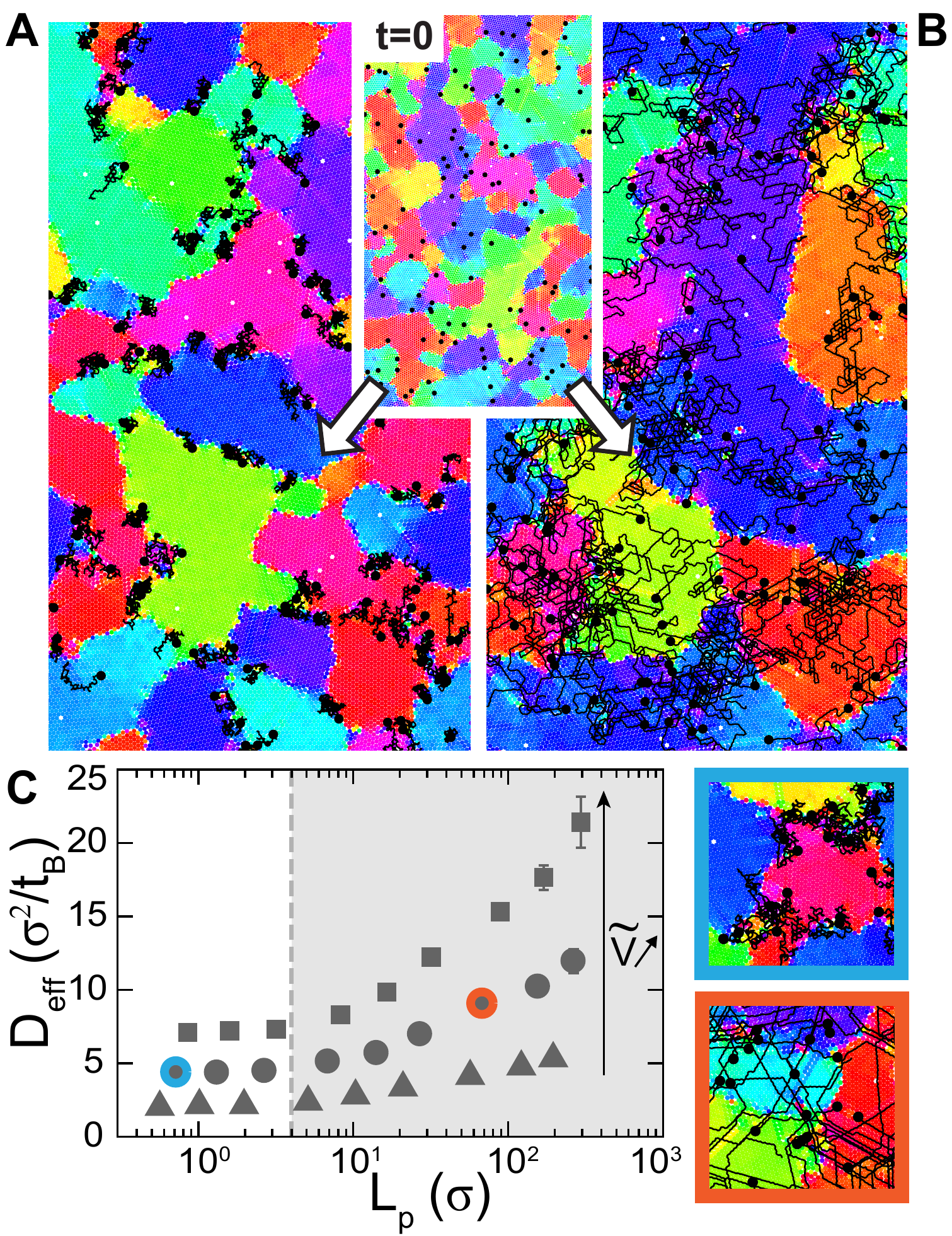}
\centering
\linespread{1}
\caption{{\bf Numerical simulation of a colloidal monolayer, with added active particles of different persistence}. Evolution of the system starting from the same initial configuration {\bf (inset, t=0)} with a fraction $\alpha=0.6\%$ of active intruders, with same speed but different persistence time, leading to persistence lengths of {\bf (A)} $L_p=0.3 \sigma$, and {\bf (B)} $L_p=7 \sigma$ comparable to experiments. The dynamics of the swimmers change qualitatively with the persistent time: they navigate along the directions of the crystal in one case (B) and  localize at the grain boundaries in the other (A). The colors correspond to the local hexactic orientational order as in [Fig.1] The active particles are shown as black points bigger than their actual size to improve visibility, and black lines trace their trajectories within the $3.10^6$ time steps preceding the snapshots. {\bf (C)}  \textcolor{black}{The diffusivity $D_{\text{eff}}$ of the passive particles shows a dynamic transition set by the persistence length $L_p$ of the active dopants and plateaus for $L_p\leq 0.4\sigma$. Active dopants co-localize at the grain boundary for $L_p\leq 0.4\sigma$ and travel in bulk otherwise, as illustrated with typical trajectories on the side panels, which correspond to the  highlighted points on the graph. The qualitative change in the annealing mechanism, as visible on panels (A) and (B), coincides with a transition in energy transfer between the dopants and the passive particles.} \label{Fig3}}
\end{figure}

\textbf{Numerical simulation of a colloidal monolayer.} Finally, we compare our experiments with the results of two-dimensional Brownian dynamics simulations\cite{anderson2008,glaser2015} of  a polycrystalline monolayer  of $2.10^4$ particles,  with $\Phi_s=0.67$, and periodic boundary conditions [Methods]. A fraction $\alpha=0.3-1.2\%$ of active particles, with diameter $0.2\sigma$,  is pulled at a constant force and displaced into the monolayer. The persistence time of the motion can be modified to explore regimes which are not accessible to the experiment. For persistence times comparable with the experiment, the simulations qualitatively capture the behavior of the system: the intruders propel, following the local order of the crystallites. They form streaks in the monolayer [Fig.\ref{Fig3}B] and speed up its annealing towards a long-range hexatic order [MovieS3]. We vary the amplitude of the pulling force and extract the velocity of the active particles,  the mobility of the grain boundary and the effective diffusivity of the passive beads as in experiments. In order to compare experimental and numerical results, we express distances in bead diameters $\sigma$ and time in diffusive time $t_B=\sigma^2/4D^*$: the time taken by a passive bead of the monolayer to diffuse one diameter $\sigma$ in the absence of activity. The numerical relationship between the resulting dimensionless velocity $\tilde V$ and effective diffusivity $\tilde D_\mathrm{eff}$  deviates from that of experiments as a result of differences in momentum transfer between both systems: absence of hydrodynamics, dissipation with the substrate, or purely two-dimensional dynamics in simulations {[SI]}. However,  the numerical result and the experiment agree reasonably when  comparing the dimensionless grain mobilities with the effective diffusivity of the passive beads, $\tilde\gamma(\tilde D_\mathrm{eff}$),  [Fig.\ref{Fig2}E], both showing an Arrhenius-like behavior. It confirms and highlights the origin of the phenomenon: active fluctuations resulting from the collision of the active beads with the passive matrix activate the process of annealing [Fig.\ref{Fig2}F]. \textcolor{black}{Larger swimmers \textcolor{black}{close in size to the passive beads, comparatively} stall in the monolayer as their motion requires the formation of energetically costly disclinations, \textcolor{black}{as previously reported by van der Meer {\it et al}\cite{van2016} in simulations where all particles have identical sizes}. Collisions with the passive matrix are largely reduced, and the effect of the active doping on the annealing \textcolor{black}{is substantially} hindered.} 
Next, we vary the persistence time of the self-propelled particles at fixed velocity in the simulations, which cannot be realized experimentally.  \textcolor{black}{\textcolor{black}{We then} observe a qualitative change in the dynamics of coarsening as a function of the \textcolor{black}{associated} persistence length \textcolor{black}{$L_p$} [Fig.3]. Swimmers with large persistence navigate through the system uniformly and form streaks \textcolor{black}{[Fig.3B]}, while self-propelled particles with short persistence co-localize at  grain boundaries [Fig.\ref{Fig3}A, MovieS4], a reminiscence of the effective attraction of hot particles in a cold bath\cite{brenner:2017fs}. The melting from the boundaries is analogous to previous numerical results\cite{kummel2015,van2016} but contrasts with our experimental observations. Further, we study the level of added fluctuations induced  by active intruders by measuring the effective diffusivity of the passive spheres \textcolor{black}{$D_\mathrm{eff}$}, as a function of \textcolor{black}{$L_p$} at fixed propulsion.  We observe a transition in  $D_\mathrm{eff}$ that plateaus for persistence lengths \textcolor{black}{shorter than the passive colloids' size}, $L_p\lesssim0.4\sigma$. The qualitative change observed in the \textcolor{black}{dopants'} dynamics [Fig. 3A and 3B]  coincides with a transition in energy transfer between the dopants and the passive particles. \textcolor{black}{The co-localization of active particles at the grain boundaries limits the collisions with the passive phase and the level of added noise: the annealing is slowed down.} \textcolor{black}{Although} the mobility of the grain is conveniently described by an activated process with an effective temperature, the observed \textcolor{black}{dynamical} transition highlights the \textcolor{black}{intrinsic} non-equilibrium nature of the annealing process.}

\begin{figure}
\includegraphics[width=0.4\linewidth]{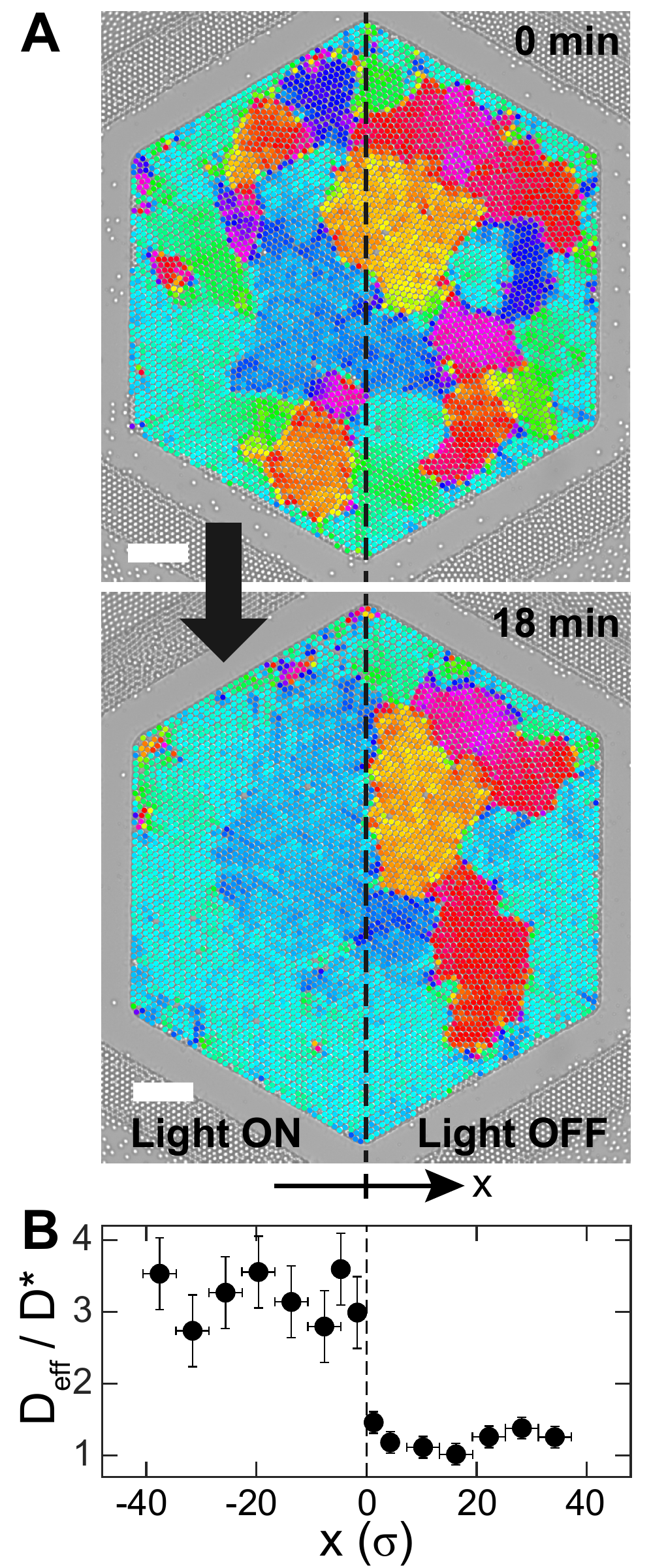}
\centering
\linespread{1}
\caption{{\bf Spatial control of the active particles' activity and effect on the monolayer.} {\bf (A)} Selective reorganization of the polycrystalline layer by activating only the intruders in the left half of the well. Scale bars, $50 \mu m$. {\bf (B)} Diffusivity of the passive particles $D_{\text{eff}}$ compared to that in a thermal system $D^*$, and plotted as a function of the distance to the illumination interface expressed in units of diameter. It illustrates dynamical control with a spatial resolution of only a few colloids. Vertical errorbars denote the standard deviation over particles located within each spatial window, whose boundaries are shown through horizontal errorbars.   \label{Fig4}}
\end{figure}

\textbf{Spatial control of the activity.} \textcolor{black}{We further exploit the potential of activity-driven assembly of colloidal materials by demonstrating spatial control of the annealing in the monolayer. Taking advantage of the  flexibility offered by photocatalytic swimmers and light patterns, we selectively activate one half of the colloidal monolayer while keeping the other half in the dark, i.e. in thermal state  [Fig.\ref{Fig4}, MovieS5].  Activity generates fluctuations in the monolayer and passive particles in illuminated regions diffuse many times faster than in dark regions, where \textcolor{black}{$D_\mathrm{eff}$} drops back to its equilibrium value [Fig.4C]. Following, the activated region quickly rearranges, while the thermal region barely evolves, illustrating a dynamical control with a spatial resolution of only a few colloids [Movie S5].  It  shows the first experimental demonstration of a man-made material controlled by internal activity, regulated in time and space.}
   
\textcolor{black}{ \textbf{Conclusion and Perspective.} Heat treatments, e.g. annealing, quenching or tempering, were originally developed for metals, to alter macroscopic properties such as strength, ductility, or toughness. Annealing is achieved through cycles of high temperatures and slow cooling: high temperatures steps allow the system to escape kinetic traps, while the slow cooling relaxes the system to its ground state with reduced defects.  We showed that active particles \textcolor{black}{embedded} in an otherwise passive system offer an  untapped approach to annealing. The addition of \textcolor{black}{\textit{in situ} large fluctuations}, mimics macroscopic annealing by heating and locally reorganizing the system. It allows local equilibration and avoids metastable states, a crucial problem in the experimental assembly of complex architectures. We further unveil a transition in the dynamics of annealing, set by the persistence of the active dopants, which has no equilibrium counterpart and had not been predicted in previous numerical works. By using light patterns, we  demonstrate a spatial control of the annealing with a spatial resolution of a few microns. The spatiotemporal control of internal fluctuations paves the way to the tuning of material properties, as well as the engineering of bio-inspired sensors \textcolor{black}{harnessing active noise for enhanced accuracy}\cite{Lan:2012in,Berut:2018iy}. \textcolor{black}{The variety of available synthetic and biological microswimmers, including particles which could remotely be steered by magnetic fields or light patterns\cite{Frangipane:2018cn,Arlt:2018cz}, offers opportunities for a wide range of control.  As grain boundaries  are essential to material properties, from yield strength\cite{Meyers:2006co,Huang:2011ew} to electrical conductivity\cite{Kreuer:2003gr},} our findings usher materials science to a new age, where the properties of matter are not controlled macroscopically but microscopically and in real time by active dopants. Further, the use of traveling particles that navigate a matrix as in this work, may suggest innovative numerical protocols of simulated annealing using virtual active intruders rather than global temperature cycling to explore complex energy landscapes.}

\section*{Material and Methods}
\textbf{Experimental set-up.} Our colloidal model system consists of silica beads of diameter $\sigma= 5\mu m$ (Sigma aldrich, 44054) suspended in a $6\%$ solution of hydrogen peroxide $H_2O_2$ in deionized water (Millipore, $18.2M\Omega$). Within a few minutes, the heavy particles sediment on the bottom wall of the sample cell to form a dense monolayer of area fraction $\Phi_S = \pi N \sigma^2/4A \approx 0.68 \pm 0.03$, where $N$ is the number of particle contained in the hexagonal area $A$. The equilibrium gravitational height is much smaller than $\sigma$, so that out-of-plane thermal fluctuations are negligible in the absence of swimmers and the system is quasi two-dimensional. When active intruders are introduced in the layer however, passive particle are observed to slightly lift from the bottom surface as smaller swimmers pass by. The numeric fraction of intruders $\alpha=N_s/N$, with $N_s$ the number of swimmers, is varied. They are light activated $2 \mu$m-diameter particles, consisting of a hematite cube embedded in a polymer bead \cite{palacci2013,palacci2014}. Under UV-light, the photo-catalytic hematite triggers the local decomposition of the hydrogen peroxide contained in the solution, creating a gradient that sets the swimmer into motion through phoretic effects. They then exhibit a persistent random walk along the bottom surface.\\The sample cell containing the solution is assembled from a glass microscope slide on top and a 0.38mm-thick polymethylmethacrylate PMMA sheet (Goodfellow ME303001) on the bottom, separated by rectangular capillaries (Vitrocom 3524) used as spacers of approximate height $600\mu$m. Shallow $4 \mu$m-deep hexagonal wells are heat embossed in the PMMA bottom surface using a polydimethylsiloxane PDMS mold fabricated via soft lithography. This confinement arena allows to keep the surface fraction $\Phi_S$ constant for the duration of an experiment. A narrow trench additionally surrounds the hexagon, to trap exterior colloids and prevent them from falling into the well. Quantitative measurements are conducted in $400 \mu$m-wide hexagons, containing about 5600 passive particles. Prior to assembly, all components are washed with Hellmanex III and thoroughly rinsed with deionized water (Milli-Q, resistivity 18.2 M). After injecting the colloidal solution, the cell is sealed with capillary wax (Hampton Research HR4-328).\\
The system is observed using an inverted optical microscope (Nikon Eclipse-Ti) equipped with a 20$\times$ objective. A LED with a wavelength $\lambda=390-480 nm$ (Lumencor, Spectra X) uniformly illuminates the hexagonal chamber through the bottom wall and activates the swimmers. The intensity of the LED can be adjusted to modify the speed of the intruders. The UV illumination is periodically interrupted for short intervals of 20s every 80s to allow for swimmers that are wedged between passive particles and the substrate to reorient through Brownian motion and escape. The evolution of the monolayer is monitored with a camera (Hamamatsu, C11440-22CU) recording images at a frame rate of 2.5fps for 90 min, and 0.1fps for 12 hours for a thermal system (with no swimmers). The position of the passive particles is extracted in each frame using standard image analysis routines\cite{blair}. \\
\textbf{Characterization of the grain structure.} The organization of the polycrystalline layer at a time $t$ is visualized and quantified through the local orientational bond order parameter $\psi_6$. For each passive particle, $\psi_6( \vec{r}_j,t)=\frac{1}{N_j}\sum_{j=1}^{N_j}e^{6i \theta( \vec{r}_{jk})}$ is computed based on the arrangement of its $N_j$ nearest neighbors (defined using Delaunay triangulation), where $\theta( \vec{r}_{jk})$ is the angle between a particle $j$ and its neighbor $k$ with respect to a reference axis chosen here as one of the direction of the hexagonal well [see SI-Fig.1b]. The amplitude of $\psi_6$, which takes values within [0.9-1] at the exception of the grain boundaries, reflects local hexagonal order [see SI-Fig.1c]. Its phase provides the local crystalline orientation $\theta_6( \vec{r}_j,t)=\frac{1}{6} \text{arg}(\psi_6( \vec{r}_j,t))$ that varies from 0 to $60^{o}$ due to rotational symmetry. In Fig.\ref{Fig1}, Fig.\ref{Fig3} and Fig.\ref{Fig4}, particles are color-coded with the crystalline orientation to visualize the grain structure.
We monitor the evolution of spatial structures using the bond-orientational correlation function $g_6(r,t)$, calculated from the six-fold bond-order parameter. As previously introduced with thermal systems\cite{lavergne2017}, the field $\psi_6$ is first smoothed by averaging local values $\psi_6(\vec{r}_j,t)$ over the two first shells of neighbors surrounding particle $j$. The resulting field is then normalized, both operations allowing for a more accurate probing of the correlations at short distances \cite{blundell1994, sicilia2008}. The bond-orientational correlation function is then computed as $g_6(r,t)=\operatorname{Re} \left( \left< \hat{\psi}_6^*(\vec{r}+\vec{r}_0,t) \hat{\psi}_6(\vec{r}_0,t) \right> \right),$ with $\hat{\psi}_6$ the smoothed and normalized parameter and $\left<.\right>$ referring to the average over all pairs of particles
separated by a distance r. This definition of $g_6(r,t)$ evaluates the spatial correlation in the argument of the bond-orientation parameter, singling out the angle component of the long-range order developing in the system over time.\\

\textbf{Numerical simulations.}   We performed Brownian dynamics simulations using HOOMD-blue \cite{anderson2008,glaser2015} to model the experiment. A two-dimensional rectangular box with periodic boundary conditions is filled with 20 000 spheres of diameter $\sigma = 1$ at an area fraction $\Phi_S = 0.05$. A fraction $\alpha$ of those particles is randomly selected to form the active particles subset. Their diameter is set to $\sigma_S = 0.2$. \textcolor{black}{All the particles interact through a purely repulsive WCA potential given by :
 \begin{eqnarray}
 & U(r) =& 4\epsilon \left[\left(\frac{\xi}{r}\right)^{2n}-\left(\frac{\xi}{r}\right)^{n}+1/4 \right] \ \text{for} \ r\le 2^{1/n}\xi, \nonumber \\
 & U(r)=& 0 \ \text{for} \ r>2^{1/n}\xi, \nonumber
 \end{eqnarray}
 where $n=6$, $\epsilon=10$ and $\xi=R_i+R_j$, with $R_{i,j}$ the radius of the interacting particles. $R_{i,j}=\sigma/2$ for passive particles and $R_{i,j}=\sigma_S/2$ for active particles.}
 
 The initial configurations are prepared by compressing the diluted box until an area fraction $\Phi_S = 0.67$ in $4.10^5$ steps of length $\Delta t=0.0001$ using the Langevin dynamics integrator. At all times, the box dimensions are set to accommodate a hexagonal lattice (i.e. $L_x = L_y\times\sqrt{3} $). During this preparation step, the temperature is linearly decreased from $kT = 2$ to $0.0005$. Once quenched, the particles form a polycrystalline 2D layer. 
 
 Thermal annealing runs are performed by letting the system relax for 500 million steps without adding activity to the small particles. For boosted annealing runs, an active force is applied to the small particles. It is parametrized by its amplitude $F_a$ and a rotational diffusion constant $D_r$ that controls the random change of direction of the active force. Typical values for $F_a$ are between 0 and 4 and shown here in the range [0-1.5], and $D_r$ is taken between 0.1 and 5. Simulations of activated monolayers are run for 100 million steps.
 \textcolor{black}{The equations of motion for the simulations are given by : 
 \begin{eqnarray}
 & m \dfrac{d\vec{v}}{dt} =& \vec{F}_\mathrm{C} - \zeta \cdot \vec{v} + \vec{F}_\mathrm{R} \nonumber \\
 & \langle \vec{F}_\mathrm{R} \rangle =& 0 \nonumber \\
 & \langle |\vec{F}_\mathrm{R}|^2 \rangle =& 2 d kT \zeta / \delta t \nonumber
 \end{eqnarray}  
 where $\vec{F}_\mathrm{C}$ is the force applied on the particles originated from all potentials and constraint forces, $\zeta$ is the drag coefficient ($\zeta=1$ here), $\vec{v}$ is the particle's velocity, $\vec{F}_\mathrm{R}$ is a uniform random force and $d$ is the dimensionality of the system ($d=2$ here).}
 
\textcolor{black}{For active particles, an active force is added to the equation of motion such that $\delta \vec{r}_i = \delta t F_a \vec{p}_i$, where $F_a$ is the active velocity and $\vec{p}_i = (\cos \theta_i, \sin \theta_i)$ is the active force vector for the particle i. The rotational diffusion of this active force vector follows $\delta \theta / \delta t = \sqrt{2 D_r / \delta t} \Gamma$, where $D_r$ is the rotational diffusion constant and the gamma function $\Gamma$ is a unit-variance random variable that decorrelates space, time and particles.}

The particles positions are recorded every 10000 steps. Following the same treatment as the experiments, the dynamic of the passive colloids and the swimmers is analyzed. For each set of parameters, six independent runs are performed to improve statistics.

\section*{REFERENCES}
\bibliographystyle{naturemag}

\begin{thebibliography}{10}
\expandafter\ifx\csname url\endcsname\relax
  \def\url#1{\texttt{#1}}\fi
\expandafter\ifx\csname urlprefix\endcsname\relax\def\urlprefix{URL }\fi
\providecommand{\bibinfo}[2]{#2}
\providecommand{\eprint}[2][]{\url{#2}}

\bibitem{Jean-perrin}
\bibinfo{author}{Perrin, J.}
\newblock \bibinfo{title}{{Mouvement brownien et r{\'e}alit{\'e}
  mol{\'e}culaire}}.
\newblock \emph{\bibinfo{journal}{Annales de Chimie et de Physique}}
  (\bibinfo{year}{1909}).

\bibitem{Brown:1828jv}
\bibinfo{author}{Brown, R.}
\newblock \bibinfo{title}{{On the particles contained in the pollen of plants;
  and on the general existence of active molecules in organic and inorganic
  bodies}}.
\newblock \emph{\bibinfo{journal}{Edinburgh new Philosophical Journal}}
  (\bibinfo{year}{1828}).

\bibitem{Einstein:1905ty}
\bibinfo{author}{Einstein, A.}
\newblock \bibinfo{title}{{The motion of elements suspended in static liquids
  as claimed in the molecular kinetic theory of heat}}.
\newblock \emph{\bibinfo{journal}{Annalen der physik}}
  \textbf{\bibinfo{volume}{17}}, \bibinfo{pages}{549--560}
  (\bibinfo{year}{1905}).

\bibitem{Brugues:2014vc}
\bibinfo{author}{Brugu{\'e}s, J.} \& \bibinfo{author}{Needleman, D.}
\newblock \bibinfo{title}{{Physical basis of spindle self-organization}}.
\newblock \emph{\bibinfo{journal}{Proceedings of the National Academy of
  Sciences of the U.S.A}} \textbf{\bibinfo{volume}{111}},
  \bibinfo{pages}{18496--18500} (\bibinfo{year}{2014}).

\bibitem{Wittmann:2001ed}
\bibinfo{author}{Wittmann, T.}, \bibinfo{author}{Hyman, A.} \&
  \bibinfo{author}{Desai, A.}
\newblock \bibinfo{title}{{The spindle: a dynamic assembly of microtubules and
  motors}}.
\newblock \emph{\bibinfo{journal}{Nature Cell Biology}}
  \textbf{\bibinfo{volume}{3}}, \bibinfo{pages}{E28--E34}
  (\bibinfo{year}{2001}).

\bibitem{Brangwynne:2008dx}
\bibinfo{author}{Brangwynne, C.~P.}, \bibinfo{author}{Koenderink, G.~H.},
  \bibinfo{author}{MacKintosh, F.~C.} \& \bibinfo{author}{Weitz, D.~A.}
\newblock \bibinfo{title}{{Cytoplasmic diffusion: molecular motors mix it up}}.
\newblock \emph{\bibinfo{journal}{The Journal of Cell Biology}}
  \textbf{\bibinfo{volume}{183}}, \bibinfo{pages}{583--587}
  (\bibinfo{year}{2008}).

\bibitem{Guo:2014dw}
\bibinfo{author}{Guo, M.} \emph{et~al.}
\newblock \bibinfo{title}{{Probing the Stochastic, Motor-Driven Properties of
  the Cytoplasm Using Force Spectrum Microscopy}}.
\newblock \emph{\bibinfo{journal}{Cell}} \textbf{\bibinfo{volume}{158}},
  \bibinfo{pages}{822--832} (\bibinfo{year}{2014}).

\bibitem{Hess:2006fq}
\bibinfo{author}{Hess, H.}
\newblock \bibinfo{title}{{Self-assembly driven by molecular motors}}.
\newblock \emph{\bibinfo{journal}{Soft Matter}} \textbf{\bibinfo{volume}{2}},
  \bibinfo{pages}{669--677} (\bibinfo{year}{2006}).

\bibitem{Vinokur:2011ub}
\bibinfo{author}{Koshelev, A.} \& \bibinfo{author}{Vinokur, V.}
\newblock \bibinfo{title}{{Dynamic Melting of the Vortex Lattice.}}
\newblock \emph{\bibinfo{journal}{Physical Review Letters}}
  \textbf{\bibinfo{volume}{73}}, \bibinfo{pages}{3580--3583}
  (\bibinfo{year}{1994}).

\bibitem{Paltiel:2000gd}
\bibinfo{author}{Paltiel, Y.} \emph{et~al.}
\newblock \bibinfo{title}{{Dynamic instabilities and memory effects in vortex
  matter}}.
\newblock \emph{\bibinfo{journal}{Nature}} \textbf{\bibinfo{volume}{403}},
  \bibinfo{pages}{398--401} (\bibinfo{year}{2000}).

\bibitem{Sengupta:2012iea}
\bibinfo{author}{Sengupta, S.}, \bibinfo{author}{Ibele, M.~E.} \&
  \bibinfo{author}{Sen, A.}
\newblock \bibinfo{title}{{Fantastic Voyage: Designing Self-Powered
  Nanorobots}}.
\newblock \emph{\bibinfo{journal}{Angewandte Chemie-International Edition In
  English}} \textbf{\bibinfo{volume}{51}}, \bibinfo{pages}{8434--8445}
  (\bibinfo{year}{2012}).

\bibitem{di2010bacterial}
\bibinfo{author}{Di~Leonardo, R.} \emph{et~al.}
\newblock \bibinfo{title}{Bacterial ratchet motors}.
\newblock \emph{\bibinfo{journal}{Proceedings of the National Academy of
  Sciences}} \textbf{\bibinfo{volume}{107}}, \bibinfo{pages}{9541--9545}
  (\bibinfo{year}{2010}).

\bibitem{sokolov2010}
\bibinfo{author}{Sokolov, A.}, \bibinfo{author}{Apodaca, M.~M.},
  \bibinfo{author}{Grzybowski, B.~A.} \& \bibinfo{author}{Aranson, I.~S.}
\newblock \bibinfo{title}{Swimming bacteria power microscopic gears}.
\newblock \emph{\bibinfo{journal}{Proceedings of the National Academy of
  Sciences}} \textbf{\bibinfo{volume}{107}}, \bibinfo{pages}{969--974}
  (\bibinfo{year}{2010}).

\bibitem{bricard2013}
\bibinfo{author}{Bricard, A.}, \bibinfo{author}{Caussin, J.-B.},
  \bibinfo{author}{Desreumaux, N.}, \bibinfo{author}{Dauchot, O.} \&
  \bibinfo{author}{Bartolo, D.}
\newblock \bibinfo{title}{Emergence of macroscopic directed motion in
  populations of motile colloids}.
\newblock \emph{\bibinfo{journal}{Nature}} \textbf{\bibinfo{volume}{503}},
  \bibinfo{pages}{95} (\bibinfo{year}{2013}).

\bibitem{Wu:2017ct}
\bibinfo{author}{Wu, K.-T.} \emph{et~al.}
\newblock \bibinfo{title}{{Transition from turbulent to coherent flows in
  confined three-dimensional active fluids}}.
\newblock \emph{\bibinfo{journal}{Science}} \textbf{\bibinfo{volume}{355}},
  \bibinfo{pages}{eaal1979--18} (\bibinfo{year}{2017}).

\bibitem{Lushi:2014fn}
\bibinfo{author}{Lushi, E.}, \bibinfo{author}{Wioland, H.} \&
  \bibinfo{author}{Goldstein, R.~E.}
\newblock \bibinfo{title}{{Fluid flows created by swimming bacteria drive
  self-organization in confined suspensions}}.
\newblock \emph{\bibinfo{journal}{Proceedings of the National Academy of
  Sciences of the U.S.A}} \textbf{\bibinfo{volume}{111}},
  \bibinfo{pages}{9733--9738} (\bibinfo{year}{2014}).

\bibitem{Cates:2015ft}
\bibinfo{author}{Cates, M.~E.} \& \bibinfo{author}{Tailleur, J.}
\newblock \bibinfo{title}{{Motility-Induced Phase Separation}}.
\newblock \emph{\bibinfo{journal}{Annual Review of Condensed Matter Physics}}
  \textbf{\bibinfo{volume}{6}}, \bibinfo{pages}{219--244}
  (\bibinfo{year}{2015}).

\bibitem{theurkauff2012}
\bibinfo{author}{Theurkauff, I.}, \bibinfo{author}{Cottin-Bizonne, C.},
  \bibinfo{author}{Palacci, J.}, \bibinfo{author}{Ybert, C.} \&
  \bibinfo{author}{Bocquet, L.}
\newblock \bibinfo{title}{Dynamic clustering in active colloidal suspensions
  with chemical signaling}.
\newblock \emph{\bibinfo{journal}{Physical Review Letters}}
  \textbf{\bibinfo{volume}{108}}, \bibinfo{pages}{268303}
  (\bibinfo{year}{2012}).

\bibitem{buttinoni2013}
\bibinfo{author}{Buttinoni, I.} \emph{et~al.}
\newblock \bibinfo{title}{Dynamical clustering and phase separation in
  suspensions of self-propelled colloidal particles}.
\newblock \emph{\bibinfo{journal}{Physical Review Letters}}
  \textbf{\bibinfo{volume}{110}}, \bibinfo{pages}{238301}
  (\bibinfo{year}{2013}).

\bibitem{palacci2013}
\bibinfo{author}{Palacci, J.}, \bibinfo{author}{Sacanna, S.},
  \bibinfo{author}{Steinberg, A.~P.}, \bibinfo{author}{Pine, D.~J.} \&
  \bibinfo{author}{Chaikin, P.~M.}
\newblock \bibinfo{title}{Living crystals of light-activated colloidal
  surfers}.
\newblock \emph{\bibinfo{journal}{Science}} \bibinfo{pages}{1230020}
  (\bibinfo{year}{2013}).

\bibitem{wu2000}
\bibinfo{author}{Wu, X.-L.} \& \bibinfo{author}{Libchaber, A.}
\newblock \bibinfo{title}{Particle diffusion in a quasi-two-dimensional
  bacterial bath}.
\newblock \emph{\bibinfo{journal}{Physical Review Letters}}
  \textbf{\bibinfo{volume}{84}}, \bibinfo{pages}{3017} (\bibinfo{year}{2000}).

\bibitem{leptos2009}
\bibinfo{author}{Leptos, K.~C.}, \bibinfo{author}{Guasto, J.~S.},
  \bibinfo{author}{Gollub, J.~P.}, \bibinfo{author}{Pesci, A.~I.} \&
  \bibinfo{author}{Goldstein, R.~E.}
\newblock \bibinfo{title}{Dynamics of enhanced tracer diffusion in suspensions
  of swimming eukaryotic microorganisms}.
\newblock \emph{\bibinfo{journal}{Physical Review Letters}}
  \textbf{\bibinfo{volume}{103}}, \bibinfo{pages}{198103}
  (\bibinfo{year}{2009}).

\bibitem{mino2011}
\bibinfo{author}{Mino, G.} \emph{et~al.}
\newblock \bibinfo{title}{Enhanced diffusion due to active swimmers at a solid
  surface}.
\newblock \emph{\bibinfo{journal}{Physical Review Letters}}
  \textbf{\bibinfo{volume}{106}}, \bibinfo{pages}{048102}
  (\bibinfo{year}{2011}).

\bibitem{Reichhardt:2004dg}
\bibinfo{author}{Reichhardt, C.} \& \bibinfo{author}{Reichhardt, C. J.~O.}
\newblock \bibinfo{title}{{Local Melting and Drag for a Particle Driven through
  a Colloidal Crystal}}.
\newblock \emph{\bibinfo{journal}{Physical Review Letters}}
  \textbf{\bibinfo{volume}{92}}, \bibinfo{pages}{095504--4}
  (\bibinfo{year}{2004}).

\bibitem{van2016}
\bibinfo{author}{van~der Meer, B.}, \bibinfo{author}{Filion, L.} \&
  \bibinfo{author}{Dijkstra, M.}
\newblock \bibinfo{title}{Fabricating large two-dimensional single colloidal
  crystals by doping with active particles}.
\newblock \emph{\bibinfo{journal}{Soft Matter}} \textbf{\bibinfo{volume}{12}},
  \bibinfo{pages}{3406--3411} (\bibinfo{year}{2016}).

\bibitem{van2017}
\bibinfo{author}{van~der Meer, B.}, \bibinfo{author}{Dijkstra, M.} \&
  \bibinfo{author}{Filion, L.}
\newblock \bibinfo{title}{Removing grain boundaries from three-dimensional
  colloidal crystals using active dopants}.
\newblock \emph{\bibinfo{journal}{Soft Matter}} \textbf{\bibinfo{volume}{12}},
  \bibinfo{pages}{5630--5635} (\bibinfo{year}{2016}).

\bibitem{kummel2015}
\bibinfo{author}{K{\"u}mmel, F.}, \bibinfo{author}{Shabestari, P.},
  \bibinfo{author}{Lozano, C.}, \bibinfo{author}{Volpe, G.} \&
  \bibinfo{author}{Bechinger, C.}
\newblock \bibinfo{title}{Formation, compression and surface melting of
  colloidal clusters by active particles}.
\newblock \emph{\bibinfo{journal}{Soft Matter}} \textbf{\bibinfo{volume}{11}},
  \bibinfo{pages}{6187--6191} (\bibinfo{year}{2015}).

\bibitem{KOmar:2018ho}
\bibinfo{author}{K~Omar, A.}, \bibinfo{author}{Wu, Y.}, \bibinfo{author}{Wang,
  Z.-G.} \& \bibinfo{author}{F~Brady, J.}
\newblock \bibinfo{title}{{Swimming to Stability: Structural and Dynamical
  Control via Active Doping}}.
\newblock \emph{\bibinfo{journal}{Acs Nano}} \textbf{\bibinfo{volume}{13}},
  \bibinfo{pages}{1--13} (\bibinfo{year}{2018}).

\bibitem{Dietrich:2018hp}
\bibinfo{author}{Dietrich, K.} \emph{et~al.}
\newblock \bibinfo{title}{{Active Atoms and Interstitials in Two-Dimensional
  Colloidal Crystals}}.
\newblock \emph{\bibinfo{journal}{Physical Review Letters}}
  \textbf{\bibinfo{volume}{120}}, \bibinfo{pages}{268004}
  (\bibinfo{year}{2018}).

\bibitem{howse2007}
\bibinfo{author}{Howse, J.~R.} \emph{et~al.}
\newblock \bibinfo{title}{Self-motile colloidal particles: from directed
  propulsion to random walk}.
\newblock \emph{\bibinfo{journal}{Physical Review Letters}}
  \textbf{\bibinfo{volume}{99}}, \bibinfo{pages}{048102}
  (\bibinfo{year}{2007}).

\bibitem{koumakis2013}
\bibinfo{author}{Koumakis, N.}, \bibinfo{author}{Lepore, A.},
  \bibinfo{author}{Maggi, C.} \& \bibinfo{author}{Di~Leonardo, R.}
\newblock \bibinfo{title}{Targeted delivery of colloids by swimming bacteria}.
\newblock \emph{\bibinfo{journal}{Nature Communications}}
  \textbf{\bibinfo{volume}{4}}, \bibinfo{pages}{2588} (\bibinfo{year}{2013}).

\bibitem{lavergne2017}
\bibinfo{author}{Lavergne, F.~A.}, \bibinfo{author}{Aarts, D. G. A.~L.} \&
  \bibinfo{author}{Dullens, R. P.~A.}
\newblock \bibinfo{title}{Anomalous grain growth in a polycrystalline monolayer
  of colloidal hard spheres}.
\newblock \emph{\bibinfo{journal}{Physical Review X}}
  \textbf{\bibinfo{volume}{7}}, \bibinfo{pages}{041064} (\bibinfo{year}{2017}).

\bibitem{sicilia2008}
\bibinfo{author}{Sicilia, J.~J., A.and~Arenzon} \emph{et~al.}
\newblock \bibinfo{title}{Experimental test of curvature-driven dynamics in the
  phase ordering of a two dimensional liquid crystal}.
\newblock \emph{\bibinfo{journal}{Physical Review Letters}}
  \textbf{\bibinfo{volume}{101}}, \bibinfo{pages}{197801}
  (\bibinfo{year}{2008}).

\bibitem{rollett2004}
\bibinfo{author}{Rollett, A.}, \bibinfo{author}{Humphreys, F.~J.},
  \bibinfo{author}{Rohrer, G.~S.} \& \bibinfo{author}{Hatherly, M.}
\newblock \emph{\bibinfo{title}{Recrystallization and related annealing
  phenomena}} (\bibinfo{publisher}{Elsevier}, \bibinfo{year}{2004}).

\bibitem{mccandlish2012}
\bibinfo{author}{McCandlish, S.~R.}, \bibinfo{author}{Baskaran, A.} \&
  \bibinfo{author}{Hagan, M.~F.}
\newblock \bibinfo{title}{Spontaneous segregation of self-propelled particles
  with different motilities}.
\newblock \emph{\bibinfo{journal}{Soft Matter}} \textbf{\bibinfo{volume}{8}},
  \bibinfo{pages}{2527--2534} (\bibinfo{year}{2012}).

\bibitem{Cates:2010um}
\bibinfo{author}{Cates, M.~E.}, \bibinfo{author}{Marenduzzo, D.},
  \bibinfo{author}{Pagonabarraga, I.} \& \bibinfo{author}{Tailleur, J.}
\newblock \bibinfo{title}{{Arrested phase separation in reproducing bacteria
  creates a generic route to pattern formation}}.
\newblock \emph{\bibinfo{journal}{Proceedings of the National Academy of
  Sciences of the U.S.A}} \textbf{\bibinfo{volume}{107}},
  \bibinfo{pages}{11715--11720} (\bibinfo{year}{2010}).

\bibitem{barrat2003}
\bibinfo{author}{Barrat, J.-L.} \& \bibinfo{author}{Hansen, J.-P.}
\newblock \emph{\bibinfo{title}{Basic concepts for simple and complex liquids}}
  (\bibinfo{publisher}{Cambridge University Press}, \bibinfo{year}{2003}).

\bibitem{anderson2008}
\bibinfo{author}{Anderson, J.~A.}, \bibinfo{author}{Lorenz, C.~D.} \&
  \bibinfo{author}{Travesset, A.}
\newblock \bibinfo{title}{General purpose molecular dynamics simulations fully
  implemented on graphics processing units}.
\newblock \emph{\bibinfo{journal}{Journal of Computational Physics}}
  \textbf{\bibinfo{volume}{227}}, \bibinfo{pages}{5342--5359}
  (\bibinfo{year}{2008}).

\bibitem{glaser2015}
\bibinfo{author}{Glaser, J.} \emph{et~al.}
\newblock \bibinfo{title}{Strong scaling of general-purpose molecular dynamics
  simulations on gpus}.
\newblock \emph{\bibinfo{journal}{Computer Physics Communications}}
  \textbf{\bibinfo{volume}{192}}, \bibinfo{pages}{97--107}
  (\bibinfo{year}{2015}).

\bibitem{brenner:2017fs}
\bibinfo{author}{Tanaka, H.} \& \bibinfo{author}{Brenner, M.}
\newblock \bibinfo{title}{{Hot particles attract in a cold bath}}.
\newblock \emph{\bibinfo{journal}{Physical Review Fluids}}
  \textbf{\bibinfo{volume}{2}}, \bibinfo{pages}{043103} (\bibinfo{year}{2017}).

\bibitem{Lan:2012in}
\bibinfo{author}{Lan, G.}, \bibinfo{author}{Sartori, P.},
  \bibinfo{author}{Neumann, S.}, \bibinfo{author}{Sourjik, V.} \&
  \bibinfo{author}{Tu, Y.}
\newblock \bibinfo{title}{{The energy--speed--accuracy trade-off in
  sensory~adaptation}}.
\newblock \emph{\bibinfo{journal}{Nature Physics}}
  \textbf{\bibinfo{volume}{8}}, \bibinfo{pages}{422--428}
  (\bibinfo{year}{2012}).

\bibitem{Berut:2018iy}
\bibinfo{author}{Berut, A.} \emph{et~al.}
\newblock \bibinfo{title}{{Gravisensors in plant cells behave like an active
  granular liquid}}.
\newblock \emph{\bibinfo{journal}{Proceedings Of The National Academy Of
  Sciences Of The United States Of America}} \textbf{\bibinfo{volume}{115}},
  \bibinfo{pages}{5123--5128} (\bibinfo{year}{2018}).

\bibitem{Frangipane:2018cn}
\bibinfo{author}{Frangipane, G.} \emph{et~al.}
\newblock \bibinfo{title}{{Dynamic density shaping of photokinetic E. coli}}.
\newblock \emph{\bibinfo{journal}{eLife}} \textbf{\bibinfo{volume}{7}}
  (\bibinfo{year}{2018}).

\bibitem{Arlt:2018cz}
\bibinfo{author}{Arlt, J.}, \bibinfo{author}{Martinez, V.~A.},
  \bibinfo{author}{Dawson, A.}, \bibinfo{author}{Pilizota, T.} \&
  \bibinfo{author}{Poon, W. C.~K.}
\newblock \bibinfo{title}{{Painting with light-powered bacteria}}.
\newblock \emph{\bibinfo{journal}{Nature Communications}}
  \textbf{\bibinfo{volume}{9}}, \bibinfo{pages}{1--7} (\bibinfo{year}{2018}).

\bibitem{Meyers:2006co}
\bibinfo{author}{Meyers, M.~A.}, \bibinfo{author}{Mishra, A.} \&
  \bibinfo{author}{Benson, D.~J.}
\newblock \bibinfo{title}{{Mechanical properties of nanocrystalline
  materials}}.
\newblock \emph{\bibinfo{journal}{Progress in Materials Science}}
  \textbf{\bibinfo{volume}{51}}, \bibinfo{pages}{427--556}
  (\bibinfo{year}{2006}).

\bibitem{Huang:2011ew}
\bibinfo{author}{Huang, P.~Y.} \emph{et~al.}
\newblock \bibinfo{title}{{Grains and grain boundaries in single-layer graphene
  atomic patchwork quilts}}.
\newblock \emph{\bibinfo{journal}{Nature}} \textbf{\bibinfo{volume}{469}},
  \bibinfo{pages}{389--392} (\bibinfo{year}{2011}).

\bibitem{Kreuer:2003gr}
\bibinfo{author}{Kreuer, K.~D.}
\newblock \bibinfo{title}{{Proton-Conducting Oxides}}.
\newblock \emph{\bibinfo{journal}{Annual Review of Materials Research}}
  \textbf{\bibinfo{volume}{33}}, \bibinfo{pages}{333--359}
  (\bibinfo{year}{2003}).

\bibitem{palacci2014}
\bibinfo{author}{Palacci, J.} \emph{et~al.}
\newblock \bibinfo{title}{Light-activated self-propelled colloids}.
\newblock \emph{\bibinfo{journal}{Phil. Trans. R. Soc. A}}
  \textbf{\bibinfo{volume}{372}}, \bibinfo{pages}{20130372}
  (\bibinfo{year}{2014}).

\bibitem{blair}
\bibinfo{author}{Blair, D.} \& \bibinfo{author}{Dufresne, E.}
\newblock \emph{\bibinfo{title}{Matlab particle tracking code retrieved from
  http://physics.georgetown.edu/matlab/}}.

\bibitem{blundell1994}
\bibinfo{author}{Blundell, R.~E.} \& \bibinfo{author}{Bray, A.~J.}
\newblock \bibinfo{title}{Phase-ordering dynamics of the o (n) model: Exact
  predictions and numerical results}.
\newblock \emph{\bibinfo{journal}{Physical Review E}}
  \textbf{\bibinfo{volume}{49}}, \bibinfo{pages}{4925} (\bibinfo{year}{1994}).

\end{thebibliography}

\begin{addendum}
\item[Supplementary Information] is available in the online version of the paper
 \item We thank S. Sacanna and M. Youssef for the  synthesis of the active particles. We thank L. Bocquet, P. Chaikin and S. Sacanna for  discussions. This material is based upon work supported by the National Science Foundation under Grant No. DMR-1554724. J.P. thanks the Sloan Foundation for support through grant FG-2017-9392. E.D. acknowledges support of the US National Science Foundation under Award Number DMR-1610788, as well as the NYU IT High Performance Computing resources, services, and staff expertise.
 \item[Author contribution] S.R. and J.P. designed the study; S.R. performed the experiments and analyzed the results; E.D. performed the numerical simulations; all authors contributed to interpreting the results and writing the paper.
  \item[Competing Interests] The authors declare that they have no competing financial interests.
 \item[Correspondence] Correspondence and requests for materials
should be addressed to J.P.~(email: palacci@ucsd.edu).
 \item[Data Availability Statement] The data that support the plots within this paper and other findings of this study are available from the corresponding author upon request.
\end{addendum}

\end{document}